\documentclass[aps,pra,amsmath,amssymb,twocolumn]{revtex4-1}
\usepackage[utf8]{inputenc}
\usepackage{bm}        
\usepackage{verbatim}   
\usepackage{braket}
\usepackage{afterpage}
\usepackage{amsmath}
\usepackage{nth}
\usepackage{physics}
\usepackage{graphicx}
\usepackage{xcolor}

\newcommand{\des}{\hat{a} }
\newcommand{\cre}{\ensuremath{\hat{a}^{\dagger} } }

\newcommand{\EJ}{\ensuremath{ E_{\text{J}}}}

\newcommand{\EJT}{\ensuremath{ \tilde{E}_{\text{J}}}}

\newcommand{\im}{\ensuremath{ \text{i} }}

\begin{document}

\title{Multipartite entanglement in a Josephson Junction Laser}

\author{Ben Lang} 
 \altaffiliation[Current address: ]{Faculty of Engineering, University of Nottingham, Nottingham NG7 2RD, UK.}
\author{Andrew D. Armour}%
 
\affiliation{School of Physics and Astronomy and Centre for the Mathematics and Theoretical Physics of Quantum Non-Equilibrium Systems, University of Nottingham, Nottingham NG7 2RD, UK}
\begin{abstract}
We analyse the entanglement in a model Josephson photonics system in which a dc voltage-biased Josephson junction couples a collection of cavity modes and populates them with microwave photons.  Using an approximate quadratic Hamiltonian model, we study the Gaussian entanglement that develops between the modes as the Josephson energy of the system is increased. We find that the modes in the system fall into a series of blocks, with bipartite entanglement generated between modes within a given block. Tripartite entanglement between modes within a given block is also widespread, though it is limited to certain ranges of the Josephson energy. The system could provide an alternative route to generating multimode microwave entanglement, an important resource in quantum technologies, without the need for ac excitation. 
\end{abstract}
\maketitle

\section{Introduction}

Combining superconducting cavities with a dc-biased Josephson junction (JJ) provides a highly tunable source of non-classical microwave photons. When a voltage $V$ is applied, a single Cooper-pair can generate one or more photons with a total energy up to $2eV$.\,\cite{Hofheinz_2011}. The strong non-linearities arising from the JJ and the ultra-strong couplings that can be achieved lead to squeezing and photon blockade\,\cite{Rolland_2019}, whilst also enabling multiplets of up to 6 photons to be generated by a single Cooper-pair\,\cite{Peugeot_2022,Sheng_2021, Lang_2021, Arndt_2022}. Such devices can also be harnessed to amplify the photon number of an input signal\,\cite{Hofheinz_number_amplification_2024} or to  realise quantum thermodynamic systems such as thermometers\,\cite{Hofer_2017} or heat engines \cite{Lorch_2018}. When a tunneling Cooper-pair generates photons in different modes, the simultaneous creation process can lead to entanglement. Recent experiments probed the entanglement generated in this way in two modes\,\cite{Peugeot21} and theoretical work has explored how three or more modes can be entangled by having their photon energies sum to that of a Cooper-pair\,\cite{Dambach_2017}. 

Whilst systems where the bias voltage is chosen to drive a single resonant photon creation process involving one or more microwave modes have been investigated extensively, rather less attention has been devoted to more complex cases where the existence of multiple equally spaced modes means that more than one process is resonant simultaneously\,\cite{Wood_2021}. One system that exploits simultaneous resonances is the Josephson junction laser (JJL)\,\cite{Cassidy_2017,Simon_2018,Lang_2023,Wang_2024}. The JJL contains a set of equally spaced cavity modes, $\omega_n=n\omega_1$ where $\omega_1$ is the fundamental mode and $n$ is an integer, to which a dc bias voltage is applied that is chosen to produce a Josephson frequency resonantly matching a higher harmonic: $\omega_J=2eV/\hbar=p\omega_1$, with $p$ an integer greater than one. Thus a tunneling Cooper pair can generate either one photon in the $p$-th mode or $p$-photons in the fundamental mode. The strong nonlinearity of the JJ means that at a threshold driving strength (controlled by the Josephson energy), the system undergoes a transition in which the fundamental mode becomes strongly excited, displaying laser-like behavior\,\cite{Cassidy_2017} and the discrete time-translational symmetry set by the Josephson frequency is broken\,\cite{Lang_2023}.

Above the transition, all of the cavity modes become excited in the JJL, leading to a spectrum that forms a comb with peaks spaced by the cavity free spectral range, $\omega_1$. In contrast, below the transition only those cavity modes with frequencies an integer multiple of $\omega_J$ are activated, producing a sparser frequency comb. In the terminology of quantum cascade lasers the symmetry unbroken regime gives a harmonic frequency comb, while the symmetry broken one produces a dense frequency comb\,\cite{Kazakov_21}. Very recent experimental work demonstrated the formation of the dense comb spectrum in the JJL\,\cite{Wang_2024}. The fact that only a dc bias is required makes the JJL a simple, low-power, route to frequency comb generation in the microwave regime\,\cite{Wang_2024}.

In previous work we used a classical (mean-field) approach to investigate how the time-translational symmetry breaking transition occurs in a simple model of a JJL system\,\cite{Lang_2023}. However, the couplings generated between multiple modes in the JJL suggest that it should be a very promising system to look for entanglement. Indeed, Gaussian entanglement has already been explored in experiments on JJ-cavity systems\,\cite{Wilson_2018, Hakonen_entanglement_2022,Jolin_2023}, though these used microwave tones at different frequencies to excite the modes rather than a simple dc bias. Furthermore, large-scale entanglement is well-known in frequency comb systems in the optical regime\,\cite{Roslund_2014,Gerke_2015}, has been predicted in vibration frequency combs \cite{Mehdi_mechanical_entanglement_2021}, and has also been found very recently in the microwave frequency regime as well using a bichromatically pumped superconducting circuit\,\cite{Jolin_2023}.

In this paper we go beyond mean-field descriptions, developing an approximate Gaussian description of the quantum state of the multimode JJL which enables us to explore the entanglement produced by the dc-biased JJ. We find that entanglement emerges well below the transition and characterise the patterns of bipartite and tripartite entanglement that emerge. We also find that the choice of voltage controls the nature of the entangled steady states that emerge, with high voltages producing many independent entangled mode pairs, and lower voltages instead entangling the modes into larger families.

The outline of this paper is as follows. We introduce our model of the Josephson-laser system in Sec.\ \ref{sec:model} and then in Sec.\ \ref{Quadratic_model} describe how an approximate effective Gaussian Hamiltonian can be derived by expanding around a fixed point of the system's classical dynamics. We check the validity of our approximate Gaussian approach using  a simplified three mode system in Sec.\ \ref{threemode} and then explore multimode bipartite entanglement in Sec.\ \ref{bipartite}. We investigate tripartite entanglement in Sec.\ \ref{tripartite} and then present our conclusions in Sec.\ \ref{conclusions}.  

\section{Model}
\label{sec:model}
\begin{figure}[t]
\includegraphics[scale=0.33]{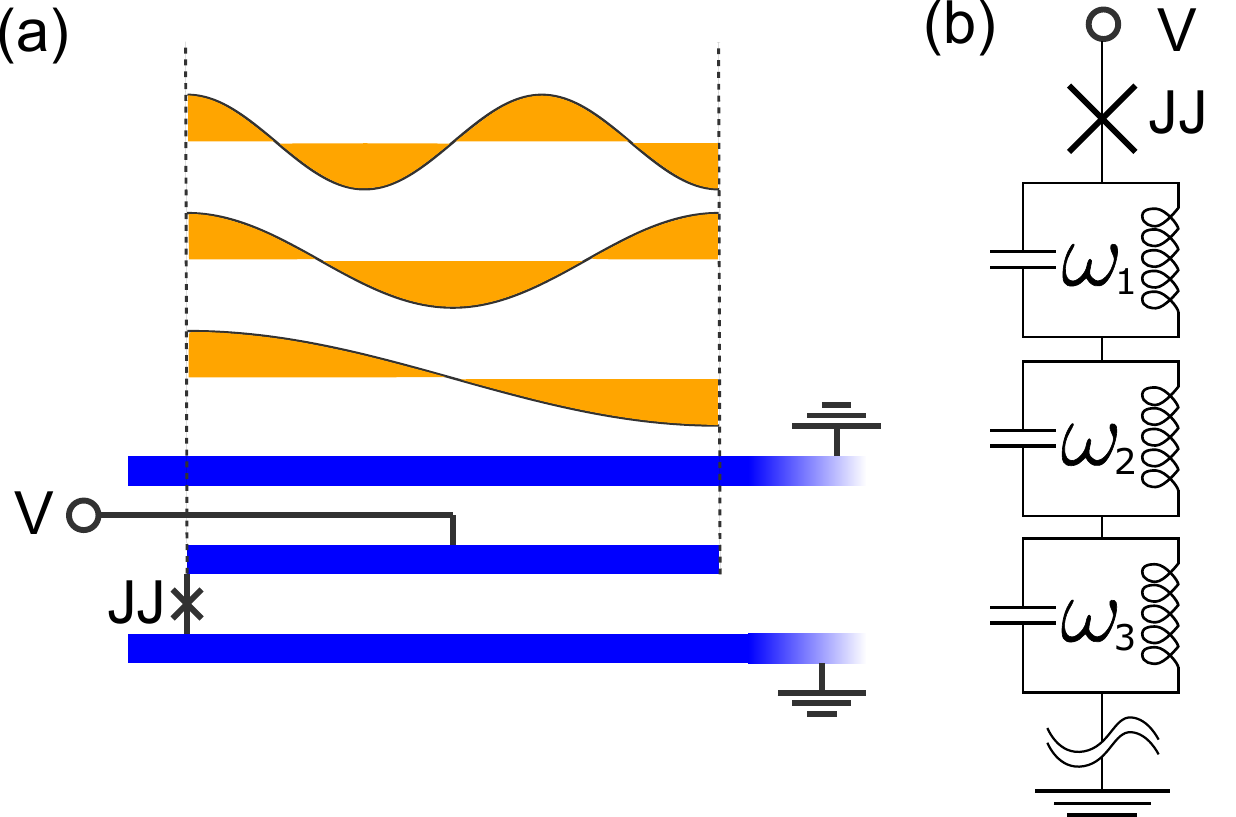}
\caption{(a) Schematic of a coplanar waveguide cavity supporting a set of harmonic modes that are equally spaced in frequency. The applied dc voltage (V) and the Josephson Junction (JJ) together produce a strongly nonlinear drive that also couples the modes. (b) Simplified circuit of the system consisting of a number of LC oscillators in series with each other and the JJ, one for each mode.}
\label{cavity_image}
\end{figure}

The system we consider is illustrated schematically in Fig.\ \ref{cavity_image}. It  consists of a microwave cavity which supports a set of harmonic modes in series with a JJ. The dc bias voltage, V,  and the JJ together lead to an effective ac drive with frequency $\omega_J = 2eV/\hbar$ which acts on the cavity modes whilst also coupling them together\,\cite{Armour_2013}. A cavity containing $N$ modes can be described by a Hamiltonian\,\cite{Armour_2013,Gramich_2013,Dambach_2017,Lang_2023}
\begin{equation}
\hat{H}(t)= \sum_{n=1}^N  \hbar \omega_{n} \cre_{n} \des_{n} - \EJ \text{cos} \left[ \omega_{\text{J}} t + \sum_{n=1}^N \Delta_{n} ( \cre_{n} + \des_{n} ) \right],
\label{time_dependent_hamiltonian}
\end{equation}
with $\EJ$ is the Josephson energy and $\des_n$ is the annihilation operator for the $n^{\text{th}}$ mode with angular frequency $\omega_n=n\omega_1$, with $\omega_1$ the fundamental, where we assume the modes are equally spaced\,\cite{Wang_2024}. The $\Delta_{n} ( \cre_{n} + \des_{n} )$ terms capture the displacements of the cavity modes with $\Delta_{n}=\sqrt{2e^2/(C\omega_n)}$ the corresponding zero-point mode displacements for a cavity with capacitance $C$\,\cite{Armour_2013}.
The leakage of photons out of the cavity modes is accounted for by using a Lindblad master equation\,\cite{Kubala_2015,Armour_2017,Dambach_2017,Lang_2021,Lang_2023}
to describe the evolution of the system's density operator
\begin{equation} 
\dot{\rho} = -\frac{\im}{\hbar}[\hat{H}(t), \rho] + \sum_{n} \frac{\gamma_{n}}{2} \mathcal{J}[ \des_n] \rho,
\label{Master_equation}
\end{equation}
where $\mathcal{J}[\des_n] \bullet = 2 \des_{n} \bullet \cre_{n}  - \cre_{n} \des_{n} \bullet - \bullet \cre_{n} \des_{n}$, $\gamma_n$ is the loss rate of mode $n$ and we have assumed that the surroundings of the cavity are at zero temperature\,\cite{Carmichael_book}.

Following \cite{Lang_2023}, we will assume a hard cutoff in the number of modes, $N$, and a constant loss rate $\gamma_n = \gamma$ \cite{Cassidy_2017,Simon_2018}. In real devices, the behavior will inevitably be more complex. Indeed, one typically expects variations in the damping rate to eventually become important as the mode index is increased\,\cite{Goppl_2008}, leading to an effective decoupling of high frequency modes, but the details of how this occurs will depend on precisely how the JJL is engineered and we will not attempt to describe it here.

We consider a voltage so that the Josephson frequency matches the $p$-th mode of the system,  $\omega_J=\omega_p+\delta=p\omega_1+\delta$, with $\delta$ a possible small detuning. This choice of Josephson frequency means that a whole range of different photon generation and exchange processes will become resonant simultaneously in the multimode system\,\cite{Wood_2021,Lang_2023}: from a process where a single photon is created in the $p$-mode at lowest order, all the way up to cases where $p$-photons are generated in the fundamental mode\,\cite{Lang_2021,Menard_2022}. Assuming that the damping is weak so that resonant processes dominate the dynamics, we obtain an effective time-independent Hamiltonian by transforming to a rotating frame using
\[
\hat{U}(t)=\exp\left(\im\sum_{n=1}^Nn(\omega_J/p)\hat{a}^{\dagger}_n\hat{a}_nt\right),
\]
and then make a RWA by neglecting terms that oscillate rapidly in time\,\cite{Armour_2013,Gramich_2013,Lang_2021,Lang_2023}. After some algebra, the Hamiltonian can be expressed in the compact form\,\cite{Lang_2023}
\begin{eqnarray}
\hat{H}_{\text{RWA}} &=& \sum_{n=1}^N \hbar \delta_{n} \cre_{n} \des_{n}-\frac{\EJT}{2} \left[  Z_{p}^{\vec{q}}(  \vec{\hat{x}} ) + {\rm{h.c.}}\right],
\label{HRWA}
\end{eqnarray}
where  $\hat{x}_n=2\im \Delta_n \des_n$,  $\vec{q}=(1,2,\ldots,N)$, $\EJT = {\EJ} \exp[-\sum_{n=1}^{N} \Delta_{n}^2 / 2]$, h.c. is  the Hermitian conjugate and $\delta_n=-(n/p)\delta$. The $Z-$functions are multidimensional functions with an appropriate number of arguments, $Z_{p}^{\vec{q}}({\vec{\hat{x}}}) = Z_{p}^{q_1, q_2, \ldots q_N}(\hat{x}_1, \hat{x}_2,\ldots\hat{x}_N)$. They provide a compact description for normally ordered power series of the mode raising and lowering operators, taking the form of analytic continuations of multidimensional Bessel functions\,\cite{Wood_2021}. The formulation in terms of $Z-$functions is useful as they are straightforward to manipulate analytically (e.g. differentiate) and they are easily evaluated numerically when they take complex number (rather than operator) arguments. Although they are described in detail in Refs.\,\cite{Wood_2021,Lang_2023}, we also provide a short summary of the properties of the $Z-$functions in Appendix \ref{App:Zfunctions} for completeness.

\section{Gaussian Effective Hamiltonian}
\label{Quadratic_model}

Previously, we analysed the mean field dynamics of the system using a coherent state ansatz in which the modes are assumed to be in coherent states described by time-dependent complex amplitudes\,\cite{Lang_2023}. This leads to an approximate description of the average amplitudes of the modes which evolve according to a set of coupled classical equations of motion. To investigate entanglement we must necessarily go beyond this simple picture, but the complex couplings between multiple modes makes a full quantum description extremely challenging, even numerically, for all but the simplest of cases. We therefore develop an effective quadratic Hamiltonian description for the system,\,\cite{Armour_2017}, obtained by expanding about a fixed point of the classical dynamics, keeping only terms of quadratic order, thereby assuming the modes are in a Gaussian state.

The quadratic Hamiltonian allows us to describe squeezing, as well as quantum correlations between the modes\,\cite{Lami_2018, Djorwe_2014}. Based on tests of the corresponding quadratic Hamiltonian for a simpler single-mode system\,\cite{Armour_2017}, we expect the Gaussian approximation to work well when the quantum fluctuations are relatively small. This requires small zero-point uncertainties for the modes\,\cite{Armour_2017} (i.e.\ $\Delta_1\ll 1$, which is the case for standard microwave cavities\,\cite{Chen_2014,Rolland_2019}), parameter regimes where the classical dynamics of the system has a single fixed point\,\cite{Wood_2021}, which is the case below the symmetry breaking transition, and that the system should not be too close to a dynamical transition (i.e.\ a bifurcation in the classical equations of motion). We further check these expectations against numerical solutions of the full quantum dynamics for a 3-mode system in Sec.\ \ref{bipartite} below.

To obtain the approximate Gaussian Hamiltonian of the system we need to expand \eqref{HRWA} about a fixed point of the classical amplitude dynamics. The fixed point amplitudes of the modes, $\vec{\alpha}=(\alpha_1,\ldots,\alpha_n)$, were analysed in\,\cite{Lang_2023}, where it was shown that they are the solutions of the set of the simultaneous equations 
\begin{equation}
\alpha_n=-\frac{2\im}{\hbar\gamma_n}\frac{\partial{\mathcal{H}}_{\rm{RWA}}(\vec{\alpha})}{\partial \alpha^*_n}, \label{classical}
\end{equation}
with $\mathcal{H}_{\rm{RWA}}(\vec{\alpha})$, the classical Hamiltonian of the system,  obtained from Eq.\ \eqref{HRWA} by replacing the raising and lowering operators by the corresponding amplitudes\,\cite{Lang_2023,Armour_2017}, e.g.\ $\hat{a}^{(\dagger)}_n\rightarrow \alpha_n^{(*)}$. 

Encoding the fixed point amplitudes in the vector, $\vec{x}_0=(2\im \Delta_1\alpha_1,\ldots,2i\Delta_N\alpha_N)$, and  expanding about the classical fixed point to second order, we obtain the Gaussian approximation to the RWA Hamiltonian 
\begin{equation}
\begin{split}
\hat{H}_{\rm{G-RWA}}(\vec{\hat{x}}) = &{\mathcal{H}}_{\rm{RWA}}(\vec{{x}}_0) + ( \mathbf{\hat{x}} - \mathbf{{x}}_0 ) \nabla {\mathcal{H}}_{\rm{RWA}}(\vec{{x}}_0) \\
&+ \frac{1}{2!} ( \mathbf{\hat{x}} - \mathbf{{x}}_0 )^T \text{\rm{Hess}}[{\mathcal{\hat{H}}}_{\rm{RWA}}(\vec{{x}}_0)] ( \mathbf{\hat{x}} - \mathbf{{x}}_0 ) \,,
\label{Taylor}
\end{split}
\end{equation}
where Hess indicates the Hessian matrix. Note that the gradient and Hessian are calculated in an expanded basis with $2N$ components, since each mode contributes two degrees of freedom\,\cite{Armour_2017}. We distinguish vectors that span all $2N$ degrees of freedom by representing them in boldface, in contrast to non-bold vectors (e.g.\ $\vec{x}_0$) which here contain $N$ components, thus, for example $\bf{\hat{x}}=(\hat{x}_1,\hat{x}^{\dagger}_1,\hat{x}_2,\hat{x}^{\dagger}_2,\ldots,\hat{x}_N,\hat{x}^{\dagger}_N)$.

We proceed by displacing to a frame centred on the classical fixed point\,\cite{Armour_2017}, by defining $\tilde{{\rho}}=\hat{D}^{\dagger}(\vec{\alpha}){\rho}\hat{D}(\vec{\alpha})$, with 
\[
\hat{D}(\vec{\alpha})=\exp \left(\sum_{n=1}^N\left(\alpha_n\hat{a}^{\dagger}_n-\alpha^*_n\hat{a}_n\right) \right).
\]
The master equation for $\tilde{{\rho}}$ has the same form as Eq.\ \ref{Master_equation}, 
\begin{equation} 
\dot{\tilde{\rho}} = -\frac{\im}{\hbar}[\hat{H}_Q, \tilde{\rho}] + \sum_{n} \frac{\gamma_{n}}{2} \mathcal{J}[\des_n] \tilde{\rho},
\label{Master_equation_Displaced}
\end{equation}
but with an effective Hamiltonian, $\hat{H}_Q$, which is now purely quadratic\,\cite{Armour_2017} (defined in Eq.\ \ref{Quad_Ham} below),
This master equation can now be used to obtain the closed set of equations of motion for the expectation values of the quadratic operators, e.g.\ $\langle \des_n \cre_m \rangle$, $\langle \des_n \des_m \rangle$. The resulting set of $2 N^2 + N$ coupled linear equations can then be solved numerically to obtain all of the second moments of the corresponding Gaussian steady state from which the squeezing and entanglement properties can then be determined as we describe below. This set of second moments fully characterise the Gaussian steady-state (since the displacement ensures that the first moments are all zero) and are conveniently collected into a covariance matrix.

For the on-resonance case (where $\delta=0$), we exploit the properties of displaced $Z-$functions (see Appendix \ref{App:displaced_z}) to  obtain
\begin{equation}
\label{Quad_Ham}
\hat{H}_{\text{Q}} = \frac{1}{2} \, (\mathbf{\cre})^T \text{diag}(\mathbf{\Delta}) \overline{M} \, \text{diag}(\mathbf{\Delta}) \mathbf{\des}
\end{equation}
with $\mathbf{\Delta} = ( \Delta_1, \Delta_1, \Delta_2, \Delta_2, \Delta_3,\ldots)$, $\mathbf{\des} = ( \des_1, \cre_1, \des_2, \cre_2,\ldots)$ and the matrix $\overline{M}$ is defined as
\begin{equation}
\begin{split}
M_{ab} =& \frac{\EJT}{2} \left[  Z_{p + c}^{\vec{q}}( \vec{x}_0 ) + Z_{p - c}^{\vec{q}}(\vec{x}_0^* ) \right], 
\label{general_m}
\end{split}
\end{equation}
where we have introduced the integer
\[
c =(-1)^{a+1} \mathbf{q}_a + (-1)^b \mathbf{q}_b \, ,
\]
using components of the vector $\mathbf{q} = (1, 1, 2, 2,\ldots n,n)$. 

Further progress can be made by focusing on a specific regime of the classical dynamics. As we showed in Ref.\ \cite{Lang_2023}, the classical dynamics displays two different regimes as a function of $\EJ$. For sufficiently low $\EJ$, there is only one stable classical fixed point. Then at a threshold value of $\EJ$, there is a bifurcation at which the discrete time-translational symmetry set by $\omega_J$ is broken\,\cite{Lang_2023} and additional fixed points emerge. Here we focus on the below-threshold regime where  the system possesses a single stable classical fixed point.

For $\EJ$ values below the threshold, all processes involve energy transfer in multiples of $\hbar \omega_J=\hbar \omega_p$. This limits which modes can interact with one another and means that only the subset of the modes with frequencies given by $m\omega_J$, with $m$ a positive integer have a nonzero mean amplitude in the steady-state\,\cite{Simon_2018,Lang_2023} (i.e.\ $\alpha_n=0$ unless $n$ is an integer multiple of $p$). This set of excited modes, which we term the  \textit{resonant harmonics}, thus form a frequency comb with spacing $\omega_J$.

In the below-threshold regime, the quadratic Hamiltonian, (\ref{Quad_Ham}), can be expressed in block diagonal form as
\begin{equation}
\hat{H}_Q = \hat{H}(0) +  \sum_{k=1}^{k_{\rm{max}}} \left[ \hat{H}(k) + \hat{H}^{\dagger}(k) \right]
\label{H_block_equ}
\end{equation}
where $k_{\rm{max}}=(p-1)/2$ for $p$ odd and $p/2$ for even $p$. The individual Hamiltonian blocks are defined by
\begin{equation}
\hat{H}(k) = \frac{1}{2} \left[\vec{s}(k)\right]^{{\dagger}T} \, \, \overline{B}(k) \, \vec{s}(k)\,,
\label{H_block}
\end{equation}
with the components of the vector $\vec{s}(k)$ alternating between annihilation and creation operators, $\des_{np+k}$ and $\cre_{np-k}$, with $n=0$, 1, 2,$\ldots$ [i.e. $\vec{s}(k) = (\des_k, \cre_{p-k}, \des_{p+k},\cre_{2p-k},\ldots)$].  The ellipsis, $\dots$, indicates that the pattern is continued outwards to include the highest mode number in the sequence that exists given the finite mode number $N$, determining the sizes of the blocks. The $k=0$ block, $\hat{H}(0)$, describes the resonant harmonics and is obtained using a  slightly modified vector of operators with a zero as the first term, $\vec{s}(0)= (0, \cre_{p}, \des_{p},\cre_{2p},\ldots)$\,\footnote{Note that when $N<p$ there are no resonant modes and $H(0)$ vanishes}. When $p$ is even, the $k=p/2$ block coupling together 'half-resonant' harmonics (i.e.\ $n=p/2,3p/2,\ldots$) has distinct dynamics due to the fact that many of the mode labels become degenerate. For example, $\des_k$ and $\cre_{p-k}$ act on the same mode.

The matrices that define the mode-mode couplings involve second derivatives of the classical Hamiltonian evaluated at the fixed point\,\cite{Armour_2017,Lang_2023} and take the form $\overline{B}(k) = \text{diag}[\vec{\Delta}(k)] \, \overline{B} \, \text{diag}[\vec{\Delta}(k)]$ with $\text{diag}[\vec{\Delta}(k)]$ the diagonal matrix with entries $( \Delta_k, \Delta_{p-k}, \Delta_{p+k},\ldots)$. Up to factors of $\Delta_n$, the blocks have a universal structure, $\overline{B}$. Defining $\vec{v} = (0, +p, -p, +2p, -2p, +3p,\ldots)$ we can express $\overline{B}_{xy} = G(v_x - v_y)$, where $G(n) = - (\EJT / 2) [  Z_{p+n}^{\vec{q}}( \vec{x}_0 ) + Z_{p-n}^{\vec{q}}( \vec{x}_0^* )] $. 

The value of $p$ (determined by the Josephson frequency, $\omega_J=p\omega_1$), controls how many non-interacting subsystems the dynamics divides into.
Within each Hamiltonian block, almost all of the information is contained in the coupling matrix $\overline{B}$. The specific blocks labelled by $k$ differ only in their sizes (set by $N$ and $p$) and by the factors $\Delta_n$. Couplings are produced between all pairs of modes within a given block: they take the form of either two-mode squeezing (producing or destroying photon pairs with frequencies that sum to an integer multiple of $\omega_J$) or beam-splitter interactions where photons are exchanged between a pair of modes with frequencies that differ by an integer multiple of $\omega_J$.

\section{Bipartite entanglement in a three mode system}
\label{threemode}
We start by exploring entanglement within a simple three-mode $(N=3)$ system where it is relatively straightforward to check our assumptions about the validity of the Gaussian analysis by comparing it with numerical calculations involving the full RWA Hamiltonian. We consider the case where $p=2$ so that for low values of $\EJ$, up until a threshold at $\EJ=E_{\text{J}}^{\rm{th}}$, only mode 2 has a non-zero fixed point amplitude, $\alpha_2$, obtained from solving Eq.\ (\ref{classical}). Above the threshold all three modes become strongly excited, breaking the time translation symmetry\,\cite{Lang_2023}. 

In the sub-threshold regime the quadratic Hamiltonian [Eq.\ (\ref{H_block_equ})] contains two blocks ($k=0,1$) and can be written as   
\begin{equation}
\begin{split}
H_Q = \frac{1}{2}  \begin{bmatrix} 
\Delta_2 \cre_2 &
\Delta_2 \des_2 
\end{bmatrix} 
\begin{bmatrix} 
g(0) & g(2)  \\
g(-2) & g(0) \\
\end{bmatrix} 
\begin{bmatrix} 
\Delta_2 \des_2 \\
\Delta_2 \cre_2 \\
\end{bmatrix}\\
+ \begin{bmatrix} \Delta_1 \cre_1 &
\Delta_1 \des_1 &
\Delta_3 \cre_3 &
\Delta_3 \des_3 &
\end{bmatrix} \\
\begin{bmatrix} g(0) & g(1) & g(-1) & g(2) \\
g(-1) & g(0) & g(-2) & g(1) \\
g(1) & g(2) & g(0) & g(3) \\
g(-2) &  g(-1)  &  g(-3) &  g(0) \\
\end{bmatrix} \begin{bmatrix} \Delta_1 \des_1 \\
\Delta_1 \cre_1 \\
\Delta_3 \des_3 \\
\Delta_3 \cre_3 \\
\end{bmatrix}
\end{split}
\label{H3}
\end{equation}
where\,\cite{Wood_2021,Lang_2023} $g(n) = G(2n) = - (\EJT / 2) [ Z_{1+n}^{(1)}( 2\im \Delta_2\alpha_2 ) + Z_{1-n}^{(1)}( -2 \im\Delta_2\alpha_2^* )]$. Both blocks in the Hamiltonian include single-mode squeezing terms (of the form $\hat{a}^{\dagger}\hat{a}^{\dagger}+\hat{a}\hat{a}$) whilst the $k=1$ block also couples modes 1 and 3. Both two-mode squeezing and beam-splitter type interactions are generated (they take the form $\hat{a}_1^{\dagger}\hat{a}_3^{\dagger}+\hat{a}_1\hat{a}_3$ and $\hat{a}_1^{\dagger}\hat{a}_3+\hat{a}_1\hat{a}_3^{\dagger}$ respectively).  In both cases, the resonant mode (mode 2) mediates the interaction by playing the role of an effective classical pump within the Gaussian approximation. As the threshold is approached from below, fluctuations in modes 1 and 3 grow, diverging at the threshold.

\begin{figure}[t]
\includegraphics[scale=0.42]{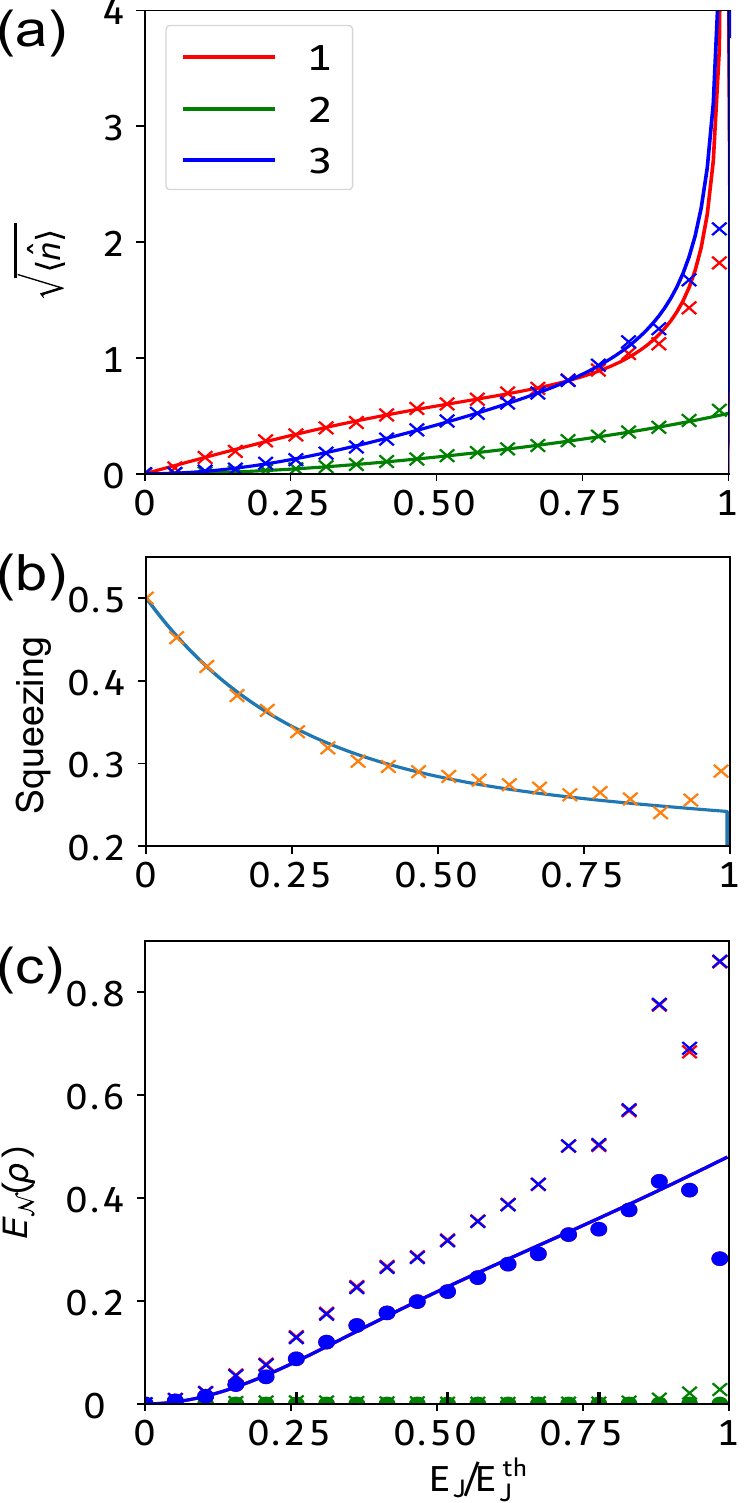}
\caption{Comparisons of the Gaussian approximation and a numerical solution using the full Hamiltonian [Eq.\ (\ref{HRWA})] for $N=3$, $p=2$. (a) Amplitude fluctuations about the fixed point in each of the three modes quantified using the corresponding occupation number in the displaced frame. The Gaussian model (lines) is compared with the numerical solutions (crosses) for modes 1 (red), 2 (green) and 3 (blue). (b) Squeezing (see the main text for a detailed definiton) from the Gaussian model (line) and numerical solution (crosses). (c) Full   ($\cross$) and Gaussian log-negativity ($\bullet$) from the numerics, with partitions separating out modes 1 (red), 2 (green) and 3 (blue). The Gaussian log-negativity for the quadratic model (line) is the same for partitions separating out either mode 1 or 3, and is always zero for the third partition. In each case $\EJ$ is scaled to the instability threshold, $E_{\text{J}}^{\rm{th}} = 1.950 \, \hbar\gamma / (\Delta_1)^2$ and a value of $\Delta_1 = 0.05$ is used throughout.}
\label{quad_p2_N3}
\end{figure}

To benchmark the Gaussian approximation, we use the  QuTip package\, \cite{QuTiP} to solve solve equation (\ref{Master_equation_Displaced}), but with an appropriately displaced version of $\hat{H}_{\text{RWA}}$ instead of the quadratic approximation, $H_Q$. Simulating three modes on a standard PC is facilitated by exploiting certain features of the $Z-$functions to save computer memory, as described in Appendix \ref{App:displaced_z}.
We truncated the Hilbert space so that the maximum number of photons in the displaced picture is 27, 5 and 25 for the three modes respectively. The steady state properties are obtained by taking an ensemble average of 1000 Monte-Carlo trajectories propagated long enough to ensure that the steady state behaviour is sampled\,\footnote{Note that as threshold is approached and fluctuations in the dynamics grow, uncertainties in the numerical results inevitably grow. This seems to affect calculations of the log-negativity in particular [see Fig. \ref{quad_p2_N3}(c)] and is reflected in the increased scatter of the points in the immediate vicinity of the threshold. However, our principal concern is with the behavior away from the immediate vicinity of threshold where there is very little scatter.}. The results from these calculations for the occupation of the modes in the displaced frame, together with the generation of squeezing and entanglement are compared to the results of the quadratic model  in Figure \ref{quad_p2_N3}. Using a low value for the zero-point fluctuations, $\Delta_1 = 0.05$, the numerical solutions allow us to test the accuracy of the Gaussian approximation away from threshold and to determine how and when it fails as the threshold is approached.

Figure \ref{quad_p2_N3}(a) shows the evolution of the fluctuations in the modes, in the form of their average occupation numbers in the displaced frame, as $\EJ$ is varied from zero up to the threshold. There is very good agreement in tracking the growth of the fluctuations as $\EJ$ increases between the numerics and the Gaussian model until near the threshold, where divergent growth in the quadratic solution for modes 1 and 3 becomes apparent, differing sharply from the much lower (finite) increase seen in the numerical solution.

Figure \ref{quad_p2_N3}(b) shows how squeezing develops as a function of $\EJ$. We plot half the smallest eigenvalue of the covariance matrix for the full 3 mode system\,\cite{Simon_1994} (see also Appendix C). A value below $1/2$ indicates squeezing, where some fluctuations are suppressed below the vacuum level. We again find good agreement, though the reduction in squeezing seen in the numerics for the immediate vicinity of threshold is not captured in the Gaussian approximation. 

Figure \ref{quad_p2_N3}(c) shows the behavior of the log negativity, $E_{\mathcal{N}}$, a well-known measure of bipartite entanglement\,\cite{Horodecki_2009}. Bipartite entanglement by definition involves partitioning the system into two subsystems which can be done in three ways for $N=3$, leading to 3 sets of crosses in Figure \ref{quad_p2_N3}(c). Also shown is the Gaussian log-negativity, calculated directly from the covariance matrix (see Appendix C and \cite{Simon_2000, Illuminati_2007}), which faithfully gives the full log-negativity only when the state is Gaussian. 

In the numerical calculation, we see a small level of entanglement between the resonant mode (mode 2) and the others, together with very slight differences between the cases where modes 1 and 3 are partitioned off. In the quadratic model the resonantly driven mode does not couple to the others [see Eq.\ (\ref{H3})] and so cannot become entangled with them. Modes 1 and 3 can only share entanglement with each other and consequently the values of $E_{\mathcal{N}}$ using the Gaussian model are the same whenever mode 1 or mode 3 is partitioned from the others.

The most striking feature of Fig.\ \ref{quad_p2_N3}(c) is that the Gaussian log-negativity is significantly below the full value, even well before the system gets close to threshold. The Gaussian model leads to a value for the log-negativity which closely matches the numerically obtained Gaussian log-negativity (except very near threshold), thus underestimating the true value of the entanglement. Thus it seems that the entanglement is sensitive to even the very small deviations from Gaussianity in the state of the system that occur well-below threshold. 

Our numerical calculations thus provide good evidence that entanglement calculated using the Gaussian model will provide an accurate estimate of the Gaussian entanglement in the full system for $\Delta_1\ll 1$, provided that one does not get too close to the threshold. They also suggest that the Gaussian model will likely act as a lower bound for the full entanglement.

\section{Bipartite entanglement in multi-mode systems}
\label{bipartite}
We now move on to use the Gaussian model to explore what happens as more modes are added in the sub-threshold regime. Within the quadratic Hamiltonian [see Eq.\ (\ref{H_block})], the modes are divided into blocks, only interacting with modes within the same block, hence entanglement is generated between modes within a given block. The number of blocks is controlled by $p$ and for $N \leq mp$, with $m=2,3,\ldots$, each subspace can contain up to $2m$ modes.

As discussed in Sec.\ \ref{Quadratic_model} above, the blocks in the quadratic Hamiltonian come in three types: one  containing the resonant harmonics ($k=0$), another coupling the half-resonant harmonics ($k=p/2$, only present for even $p$) and a set of `typical' blocks that together include all the other modes. Up to numerical factors of $\Delta_n$, any two blocks of the same type and size generate the same pattern of couplings and hence entanglement.

To understand the extent of the entanglement that is generated we thus need to examine the entanglement within each block type as a function of its size. For each block we consider all possible bipartitions of the modes.
We varied ($p,N$) in order to check typical blocks of up to 12 modes and blocks of the other two kinds with up to 5 modes. We found steady-state entanglement, $E_{\mathcal{N}} >0$,   for all possible bipartitions within a given block (for all block types) for all $\EJ > 0$, up to the threshold. This means that for any way of dividing the modes of a block into two groups, there is entanglement between the two groups, a condition known as full inseparability\,\cite{Teh_2014_genuine}. This shows that webs of entanglement span the full range of modes within each block.

\begin{figure}
\includegraphics[scale=0.5]{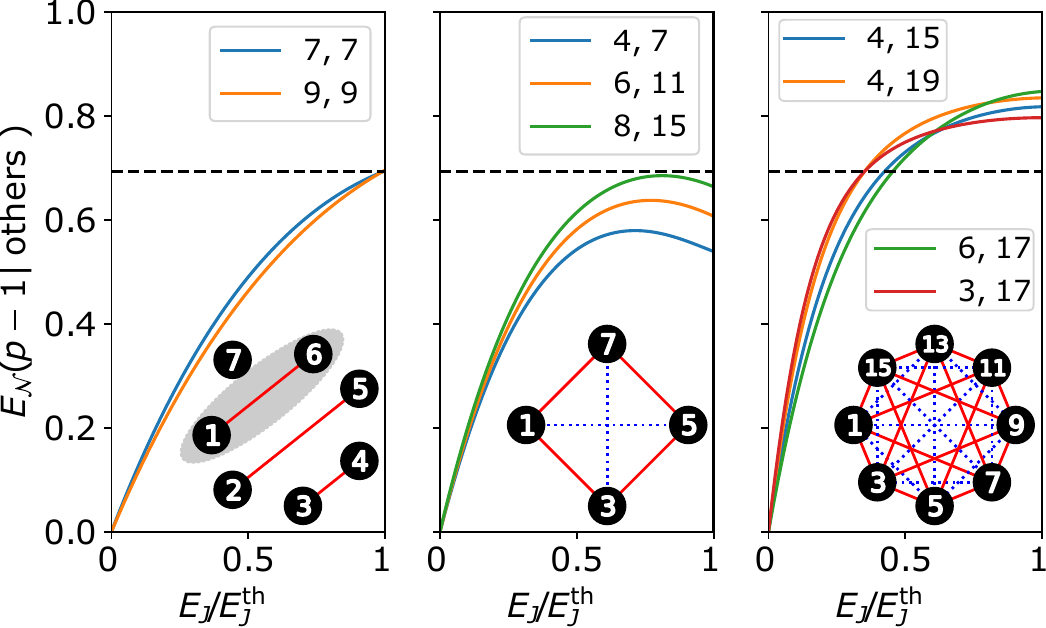}
\caption{Entanglement between mode $p-1$ and the others for (a) $N=p$. (b) $p<N<2p$. (c) $N>2p$. The legends give the $p$, $N$ values and the horizontal dashes indicate $E_{\mathcal{N}}=\ln(2)$, the maximum entanglement reached in a parametric amplifier. Insets: Schematics of the inter-mode couplings generated within Hamiltonian blocks [defined in Eq.\ (\ref{H_block_equ})] for $p,N=7,7$ (a), $4,7$ (b) and $4,15$ (c). Solid (dashed) lines indicate two-mode squeezing (beam-splitter) interactions between modes (numbered circles). In (a) the $k=1$ space is highlighted. In (b, c), only the $k=1$ space (which contains  mode $p-1$) is shown.}
\label{Entanglement_lines}
\end{figure}

Next we examine the strength of the bipartite entanglement by looking at the corresponding value of $E_{\mathcal{N}}$. In all the cases we considered we found that the entanglement turned out to be largest when  mode $p-1$ (which lies within the typical block with $k=1$) is partitioned from the others and its behavior is shown as a function of $\EJ/E_{\text{J}}^{\rm{th}}$ for a range of $(p,N)$ values in Fig.\ \ref{Entanglement_lines}. 

Figure \ref{Entanglement_lines}(a), is for the case where $N=p$ so that the typical blocks couple only 2 modes. Modes 1 and $p-1$ are coupled by a two-mode squeezing interaction and hence act as a non-degenerate parametric amplifier. As one would expect for a parametric amplifier, $E_{\mathcal{N}}$ reaches a maximum value of $\ln(2)$ at threshold\,\cite{Woolley_2014}. For $N=2p-1$, shown in Fig.\ \ref{Entanglement_lines}(b), we see more complex behaviour arising from the fact that mode $p-1$ is now  coupled to three others. Values of $E_{\mathcal{N}}>\ln(2)$ are achieved only when $N\geq 2p$, see Fig.\ \ref{Entanglement_lines}(c). We interpret this as the effect of the frequency comb: entanglement does not grow beyond the parametric limit until we have two or more resonant harmonics. Keeping $N=qp-1$ and $q>2$ the behaviour is fairly insensitive to $N$ and $p$, leading to a peak value of $E_{\mathcal{N}}\sim 0.85$.

\section{Genuine Tripartite Entanglement}
\label{tripartite}
So far we have found full inseparability in the blocks, meaning that every part of each block shares some entanglement with the rest of the block. We now look to see if the modes satisfy a more stringent condition: sharing genuine multimode entanglement. The word 'genuine' in reference to entanglement between $N$ modes means that any decomposition of the density matrix into pure states must contain at least one pure state that is $N$ mode entangled (i.e\ it cannot be written using tensor products of states that span less than $N$ modes). It stands in contrast\footnote{A simple analogy can help illustrate the distinction. Consider a set of three individuals: A, B and C. Tripartite entanglement is analogous to the situation where there is a set which includes all three individuals. In the alternative situation in which three sets exist, one including A and B, another B and C, and a third A and C, then we have something analogous to full inseparability without tripartite entanglement.} to full inseparability between N modes which can be achieved by taking a statistical mixture of quantum states each of which only has entanglement between two of the modes\,\cite{Teh_2014_genuine, Wilson_2018}.

Genuine tripartite Gaussian entanglement, which is the relevant property for a system in a Gaussian state, is detected using the method described in\,\cite{Teh_2014_genuine}. To look for entanglement involving $M$ modes we first trace out all other modes, then choose linear combinations of the mode positions and momenta: $\hat{u} = \sum_i h_i \hat{x}_i$ and $\hat{v} = \sum_i g_i \hat{p}_i$, where $h_i$ and $g_i$ are numerical factors to be varied and $\hat{x}_i=(\hat{a}_i+\hat{a}^{\dagger}_i)$, $\hat{p}_i=i(a^{\dagger}_i-a_i)$. The expected variances of $\hat{u}$ are found as $\langle \Delta u^2 \rangle = \sum_{nm} h_n h_m C_{nm}^{\text{x}}$ for $\overline{C}^{\text{x}}$ the covariance matrix of positions alone (see Appendix C), with an analogous definition for $\hat{v}$. $M$-partite entanglement is confirmed if any $\hat{u}$ and $\hat{v}$ can be found such that:
\begin{equation}
\langle \Delta u^2 \rangle + \langle \Delta v^2 \rangle < 2 \, \text{min}\{S_b\}
\label{tripartite_entangle}
\end{equation}
where the set $\{S_b\}$ has one member for each possible bipartition of the $M$ modes and for any particular bipartition
\begin{equation}
S_b = \Biggl|\sum_{\substack{\text{modes on}\\ \text{one side}}} h_i g_i \Biggr| + \Biggl|\sum_{\substack{\text{modes on}\\ \text{other side}}} h_i g_i \Biggr|.
\end{equation}
For example, for the biparition of three modes as $1|23$ $S_{1|23} = |h_1 g_1| + | h_2 g_2 + h_3 g_3|$.

We used this to explore which mode trios were entangled. If values of $\vec{h}$ and $\vec{g}$ are found that satisfy equation (\ref{tripartite_entangle}) it confirms the state contains tripartite entanglement. However, if no such vectors are found it doesn't prove the absence of entanglement.

It is easy to check if a given state and pair of vectors satisfy the entanglement condition (\ref{tripartite_entangle}). However, finding suitable vectors to reveal entanglement is harder. We used a handful of different scipy optimisation routines for searching \cite{SciPy_2020}. In the first instance the Nelder Mead method was used. If this detected no entanglement the Powel method was also attempted. We expect many of our states to be similar to one another (for example consecutive $\EJ$ values for the same system, or $N$, $p$ combinations that produce similar blocks). With this in mind before each optimisation we first tried a small library of vectors that had been found by the optimiser on previous occasions. In addition to often saving the need for further optimisation runs this also improved our sensitivity, as the optimiser would occasionally fail to rediscover previously effective vectors on its own\,\footnote{Curiously, rounding the vector elements in our library to simple fractions made them better, in that they revealed entanglement in a wider range of states than they did without rounding.}. 

\begin{figure}
\includegraphics[scale=0.6]{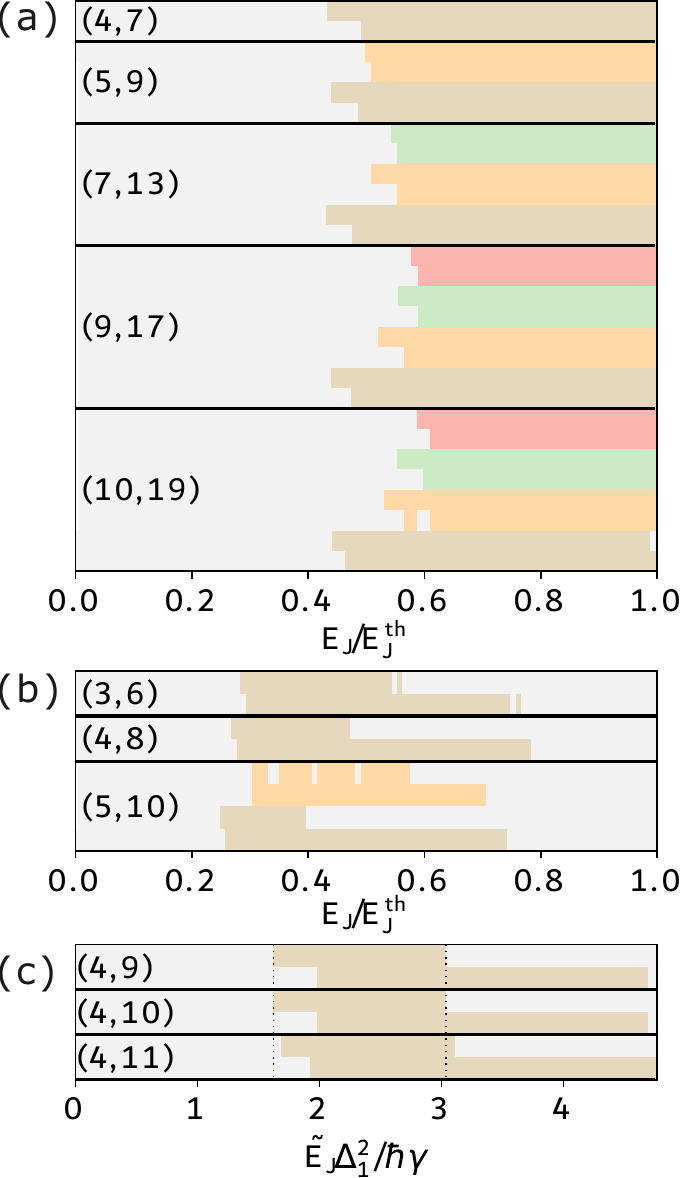}
\caption{Genuine tripartite entanglement detected (color) or not detected (gray) as a function of $\EJ$. Black horizontal lines divide the plot into sub-plots labelled by their specific $(p,N)$ values. Each $(p,N)$ sub-plot contains individual bars in pairs with  [$k$, $p-k$, $p+k$] and [$k$, $p-k$, $2p-k$] with $k$ increasing upwards. Entangled regions are coloured beige, yellow, green and red for $k=1,2,3,4$ respectively. For example the lowest horizontal bar in each sub-plot represents the mode trio [$1$, $p-1$, $p+1$ ], the next lowest [ $1$, $p-1$, $2p-1$].  (a) and (b) show tripartite entanglement as a function of $\EJ/E_{\text{J}}^{\rm{th}}$ for $N=2p-1$ and $N=2p$, respectively. The effect of adding additional modes is shown in (c), using a different $\EJ$ scale.}
\label{tripartite_ranges}
\end{figure}

In practice, we searched for tripartite entanglement within the Hamiltonian mode blocks of different types with a range of different sizes as this allowed us to identify the entanglement for all $p,N$ values for which such blocks occur. Typical blocks of all sizes up to 12 modes were sampled as a function of $\EJ$, as well as the two other block types with up to 5 modes. Overall, we were able to search for  tripartite entanglement in 38 ($p$, $N$) combinations. One pattern that was robust was that for all of the ($p$, $N$) values that we looked at, there was at least somewhere within the range $0<\EJ<E_{\text{J}}^{\rm{th}}$ where tripartite entanglement was found connecting the three modes with numbers [$k$, $p-k$, $p+k$] and the alternative set [$k$, $p-k$, $2p-k$], for all integer values $0<k<p/2$. That is, once $p$ and $N$ are fixed, there will be many different trios of modes that share tripartite entanglement. One such trio is [$1$, $p-1$, $p+1$], another [$2$, $p-2$, $p+2$] and so on. These trios always share tripartite entanglement  regardless of the number of modes $N$ or Josephson frequency $p\omega_1$, so long as the trios actually exist. However, the specific $\EJ$ ranges over which the tripartite entanglement was found depends on $N$ and $p$, as shown in Fig.\ \ref{tripartite_ranges}.

We start by surveying the case where $N=2p-1$, shown in Fig.\ \ref{tripartite_ranges}(a). For each ($p$,$N$) combination we consider the tripartite entanglement in the special mode trios discussed in the previous paragraph. As already discussed, changing $p$ while preserving the relation with $N$ (in this case $N=2p-1$) does not change the structure of the blocks, but only some of the $\Delta$ values scaling processes. This manifests as the entanglement appearing at approximately the same $\EJ/E_{\text{J}}^{\text{th}}$ value across different $p$ values for a given block number. For example entanglement appears at  
$\EJ / E_{\text{J}}^{\text{th}} \approx 0.45$ in the first block ($k=1$, beige).

Figure \ref{tripartite_ranges}(b) illustrates cases where  $N=2p$. Here the behavior changes because mode $2p$ is a resonant harmonic, populated by up-conversion from mode $p$ and has nonzero mean amplitude. The inclusion of another nonzero amplitude mode in the fixed point changes the values of the $Z-$functions appearing in the quadratic Hamiltonian in a way that suppresses fluctuations in the $k=1$ subspace and greatly delays the transition. We should distinguish two quantities, first: the lowest $\EJ$ at which time translation symmetry can be broken in the classical dynamics, and: the highest $\EJ$ at which it can remain unbroken. The two coincide in part (a) (and all previous figures). However, in \ref{tripartite_ranges}(b) the two do not coincide, and here we take $E_\text{J}^{\text{th}}$ as the former, which is the lower of the two values, which ensures that we remain in the regime with only a single stable fixed point. In this latter first-order type of transition, a new stable solution to (\ref{classical}) suddenly appears at some $\EJ$, distant in the phase space from the other (previously unique) stable solution with which it typically co-exists over a range of $\EJ$ values\,\cite{Strogatz_book}. Numerical integration is used to find these solutions, as described in\,\cite{Lang_2023}. Importantly, this type of transition does not result from a divergence in the Gaussian fluctuations, these remain finite.

In each of the pairings in Fig.\ \ref{tripartite_ranges}(b), we see the same pattern: entanglement switches on for a while before switching off again before the threshold is reached. This is most likely because the amplitude of mode $2p$ initially grows quadratically with $\EJ$, so its suppressing effect is not significant until moderate $\EJ$ values are reached. Here there is no instability, so there is no reason to expect the fluctuations relevant to entanglement to grow monotonically with $\EJ$\,\footnote{The fluctuation strengths for the individual modes measured in terms of the corresponding occupation number in the displaced frame, $\sqrt{\hat{n}}$, peak long before threshold.}. Notice that 'barcode'-like features start to appear in which there is rapid switching between regions in which entanglement is and isn't detected as a function of $\EJ$. An obvious possibility is that the optimisation procedure cannot always find an appropriate vector to reveal the entanglement, leaving small gaps.

In Fig.\ \ref{tripartite_ranges}(c) we keep $p$ fixed and add modes to the system one at a time. Going from ($p$,$N$) = (4,9) to (4,10) we see that the new mode has no impact at all on when entanglement is found in the $k=1$ block (as highlighted by the dashed lines). Indeed, it has no effect at all on the covariance matrices relating to the $k=1$ block themselves. This is simply because the newly added 10th mode isn't in the $k=1$ block (so doesn't couple to them), and also isn't in the $k=0$ block (where it would move the fixed point). In contrast, going to (4,11) we see that adding the 11th mode (which is in the $k=1$ block) does have an effect (albeit small) on the entanglement range.

Overall, the most important point is that whilst the specific ($p,N$) values change the ranges of $\EJ$ where tripartite entanglement occurs, they don't alter the general pattern that entanglement is always found {\emph{somewhere}} for the trios [$k$, $p-k$, $p+k$] and [$k$, $p-k$, $2p-k$], with $0<k<p/2$. The values of $N$ and $p$ determine how many blocks are needed to accommodate all of the modes and the sizes of the individual blocks (including the number of resonant modes).  However, the basic structure of the mode couplings within a block is encapsulated in the matrix $\overline{B}$, which is independent of the system parameters [defined in Sec.\ \ref{Quadratic_model} above] and it is presumably this feature which leads the very general pattern of entanglement that emerges.

The fact that general statements can be made about which modes are entangled for a wide range of different mode numbers and bias voltages is interesting in its own right. But it also suggests that the entanglement will be robust. Considering the trio [$1$, $p-1$, $p+1$], the parametric terms creating photon pairs in the first two and last two modes have $\Delta$-dependent pre-factors in the ratio $\sqrt{p+1}:1$, yet between ($p$,$N$) = (4,7) and (10,19) there is little change in the pattern of entanglement. Thus, if the $\Delta$ values were perturbed for any other reason (for example imperfections in fabrication) it would not have a significant effect.

Finally, we attempted to apply our methods to quadripartite entanglement, but did not find any. While it is possible that this is because none exists, it is more likely that our numerical solver failed to find appropriate vectors in the 8-dimensional search space needed for 4-partite entanglement. Numerical searches for genuine entanglement across many modes have been successful elsewhere, but made use of genetic algorithms \cite{Gerke_2015}, which are more involved than the methods applied here, which were restricted to Scipy optimisation libraries.

\section{Conclusions}
\label{conclusions}
We have explored the quantum fluctuations and entanglement properties of a model JJL system, consisting of a cavity supporting multiple modes in series with a JJ and biased by a dc voltage\,\cite{Cassidy_2017,Simon_2018,Lang_2023}. The system differs from the set-ups typically used to generate multimode entanglement in superconducting circuits\,\cite{Wilson_2018, Hakonen_entanglement_2022, Jolin_2023}, which rely on ac microwave drives and was recently used to generate a dc powered microwave frequency comb\,\cite{Wang_2024}. We found that entanglement turns out to be widespread in the system, suggesting that the JJL could be a versatile tool for generating multimode entanglement without the need for ac drives.

We used an approximate Gaussian Hamiltonian to describe the system in regimes where the Josephson frequency set by the bias voltage is resonant with one of the modes and the Josephson energy is not too high, so that the time-translational symmetry of the Hamiltonian is preserved and the system possess a single classical fixed point. The analysis is much simplified because the system can be described by a simpler effective Hamiltonian in which the all-to-all mode couplings present in the original Hamiltonian reduce to a smaller set of relevant interactions that allow the modes of the system to be split into discrete blocks (i.e.\ subspaces), with entanglement only ever developing between modes within a given block. In the most extreme case, where the Josephson frequency is resonant with the highest cavity mode, the system reduces to a set of parametric amplifiers arising within blocks that each couple a different pair of modes. In contrast, for values of the Josephson frequency resonant with lower modes, the blocks couple more modes and so can play host to more complicated multimode correlations. Investigating cases with up to 12 modes within a given subspace, we found that there is full inseparability between the modes in a given block, and in many cases genuine tripartite entanglement. 

Extensions of our model to include a soft rather than hard cutoff in the mode number as well as for scatter in the other parameter values would be numerically demanding, but otherwise straightforward. However, we would not expect such effects to lead to significant changes in the behavior as the basic patterns of entanglement were generic and seem to be robust across widely different choices of the bias voltage and the number of modes included. Another possible extension is to use the quadratic model solutions to apply squeezing operators to the full equation of motion, enabling a small Hilbert space to model the higher order quantum features that remain after the displacement and squeezing are accounted for.

\section*{Data Availability}

The simulation and plotting codes, along with the relevant data that support the findings of this article are openly available \cite{Open_Data_Lang_2024}.

\section*{Acknowledgements}

We thank Chris M. Wilson and Adam Gammon-Smith for very helpful discussions. The work was supported by a Leverhulme Trust Research Project Grant (RBG-2018-213).
\appendix
\section{$Z-$functions}
\label{App:Zfunctions}

The $Z-$functions used in this paper were introduced in \cite{Wood_2021}, based on higher dimensional Bessel functions discussed in\,\cite{Dattoli1998, Korsch2008}. We provide the key definitions here to make the text more self-contained, but refer the reader to \cite{Wood_2021} for further details.

The $Z-$functions are defined by
\begin{equation}
Z_{p}^{\vec{q}} ({\bf{\hat{x}}}) =  : \int_{-\pi}^\pi \frac{dt}{2\pi} \exp{ \sum_{l=1}^N \frac12 \left(\hat{x}_l {\rm{e}}^{\im q_l t} - {\rm{h.c.}} \right) -\im p t }:\,
\label{NDZint}
\end{equation}
with colons indicating normal ordering. The functions are easily differentiated
\begin{equation} \label{derivs}
\begin{split}
\frac{d}{d \hat{x}_j} Z_p^{\vec{q}} \left( {\bf{\hat{x}}} \right) &= \frac{1}{2}  Z_{p-q_j}^{\vec{q}} \left( {\bf{\hat{x}}} \right)
\\
\frac{d}{d \hat{x}_j^{\dagger}} Z_p^{\vec{q}} \left( {\bf{\hat{x}}} \right) &= -\frac{1}{2}  Z_{p+q_j}^{\vec{q}} \left( {\bf{\hat{x}}} \right)\,.
\end{split}
\end{equation}

\section{Displaced $Z-$functions}
\label{App:displaced_z}
To study the quantum fluctuations of the system about a fixed point it is necessary to displace the Hamiltonian. Given the Hamiltonian, equation (\ref{HRWA}), this problem reduces to finding a convenient way of displacing the $Z-$functions.

Using equation (\ref{NDZint}) a displaced $Z-$function can be expressed as a $Z-$function with additional dimensions with the frequencies of the displaced modes. For a displacement of all modes:
\begin{equation}
{D}(\vec{\alpha}) Z_p^{\vec{q}} \left( \vec{ \hat{x}} \right) =  Z_p^{\vec{q}, \vec{q}} \left( \vec{ \hat{x}}, 2 \im \vec{\Delta} \vec{\alpha} \right),
\end{equation}
where the terms following vectors with a comma are appended to those vectors and $\vec{\Delta} \vec{\alpha}$ is an element-wise product.

We then exploit an identity that enables a $Z-$function to be expressed as a sum over a product of lower-dimensional $Z-$functions\,\cite{Wood_2021}:
\begin{equation}
Z_p^{\vec{q}, \vec{q}} \left( \vec{ \hat{x}}, 2 \im \vec{\Delta} \vec{\alpha} \right) = \sum_{k=-\infty}^{\infty} Z_{p - k}^{\vec{q}} \left( \vec{ \hat{x}} \right) \, Z_{k}^{\vec{q}} \left( 2 \im \vec{\Delta} \vec{\alpha} \right).
\label{series}
\end{equation}
To derive equation (\ref{general_m}) for the quadratic Hamiltonian we use (\ref{series}) to express the displaced Hamiltonian, then differentiate the operator part to second order to find the Hessian. 
This is equivalent to identifying the quadratic terms appearing in the Taylor series of the operator part in (\ref{series}).

For numerical simulations we also make use of displaced $Z-$functions. These are handled using the series (\ref{series}) and suitably truncating the infinite sum (we find 16 terms more than sufficient). For small Hilbert spaces the operator parts of this sum can be implemented on the computer by way of filtering an exponential matrix, as described in \cite{Wood_2021}. However, for a large state space our computer does not have enough memory to exponentiate such a big matrix. To overcome this we apply the summation trick as in (\ref{series}) a further two times: so that instead of taking the exponential of matrix the size of the full Hilbert space we instead have a double sum over tensor products of smaller matrices. The terms in this double sum are found using the exponentiation, but they are only of the size of the spaces of the individual modes so require much less memory.

\section{Covariance Matrix}
\label{App:Covariance_matrix}

The covariance matrix provides a convenient way of representing a Gaussian state. This appendix summarises some relevant properties of it to make the present paper more self contained. For further details, see for example Ref.\ \cite{Illuminati_2007}.

A Gaussian quantum state is characterised by its centre of mass in phase space (represented by a displacement vector), and its spread (represented by a covariance matrix). The covariance matrix for $N$ modes is a $2N$ square matrix. Each entry is the expectation value of a symmeterized operator after the displacements are subtracted from the state. For the vector ${\bf{R}} = (\hat{x}_1, \hat{p}_1, \hat{x}_2, \hat{p}_2, \hat{x}_3, \hat{p}_3,...)$ the covariance matrix elements are given by:
\begin{equation}
C_{nm} = \langle R_n R_m + R_m R_n \rangle - 2 \langle R_n \rangle \langle R_m \rangle 
\end{equation}
Note that our $C$ differs from the $\sigma$ of \cite{Illuminati_2007} by a factor of 2, leading to similar factors differing between our equations.

For a vacuum state $\overline{C}$ is the identity matrix. All physical covariance matrices must satisfy the \textit{bona fide} condition. Defining $\omega = \begin{bmatrix} 0 & 1 \\ -1 & 0\end{bmatrix}$ and $\Omega = \omega^{\oplus N}$ ($N$ copies of $\omega$ put corner to corner to make a square matrix of size $2N$) this condition requires that all eigenvalues of $i \Omega \overline{C}$ have modulus greater than 1.

Up to displacements, $\overline{C}$ fully characterises the Gaussian state. For example, the squeezing of the state is given by half the smallest eigenvalue of $\overline{C}$.

Whether the state described by $\overline{C}$ is bipartite entangled with respect to a given bipartition can be found as follows. First, we time-reverse the modes on one side of the partition by reversing the sign of all momenta on that side \cite{Simon_2000}, which multiplies the columns and rows of $\overline{C}$ relating to those operators through by a factor of $-1$. If this partially reversed matrix, $\overline{C}^{\text{PT}}$, does not satisfy the \textit{bona fide} condition then there is entanglement connecting the parts on either side of the chosen partition. The log-negativity measures how strongly the condition is violated. The eigenvalues of $\im \Omega \overline{C}^{\text{PT}}$ are found, $\{\lambda_n \}$. The log-negativity is $-1$ times the sum of the logs of all eigenvalues between 0 and 1:
\begin{equation}
E_{\mathcal{N}} = - \sum_n \log(\lambda_n) \,\text{ for } 0 \leq \lambda_n < 1.
\end{equation}

The matrix $\overline{C}^{\text{x}}$ used in finding tripartite entanglement is simply the matrix $\overline{C}$ with the even columns and rows omitted. Similarly $\overline{C}^{\text{p}}$ omits the odd numbered rows and columns.

\bibliography{references}

\end{document}